# Oxygen Channels and Fractal Wave-Particle Duality in the Evolution of Myoglobin and Neuroglobin

Vedant. Sachdeva and James C. Phillips

Dept. of Physics and Astronomy, Rutgers University, Piscataway, N. J., 08854

Abstract

The evolution of terrestrial and aquatic wild type (WT) globins is dominated by changes in two proximate - distal Histidine ligand exit channels, here monitored quantitatively by hydropathic waves. These waves reveal allometric functional features inaccessible to single amino acid stereochemical contact models, and even very large all-atom Newtonian simulations. The evolutionary differences between these features between myoglobin and neuroglobin are related to the two oxidation channels through hydropathic wave analysis, which identifies subtle interspecies functional differences inaccessible to traditional size and metabolic scaling studies. Our analysis involves dynamic synchronization of allometric interactions across entire globins.

**Introduction**

Here we discuss the evolution of myoglobin (Mb) and neuroglobin (Nb) in functional terms, connecting hydropathic allometric globular properties and oxidation kinetics. Our analysis assumes that globins have evolved from aquatic predecessors to take nearly optimal advantage of the 28 times larger oxygen concentration available terrestrially. Mb stores $O_2$ in tissues, was the first protein whose structure was determined [1], and is the best understood globin. Nb is concentrated in the neural network and the retina, and it exhibits instructive differences from Mb, and substantial sequence differences (24% identity, 39% similarity). Features of Mb and Nb evolution are largely hidden from structural studies, which generally show little difference between human and chicken protein backbones. Phylogenetic trees explore evolution in terms of single-site amino acid (aa) identities or similarities; more complex overall evolutionary scenarios are poorly described by such one-dimensional models [2].

It was first suggested that short-range hydrogen bond donation by the neutral NϵH tautomer of distal His(E7) regulates Mb oxygen affinity, a small energy change [3]. Later structural studies showed that ligand entry into Mb occurs by swinging the distal His E7 imidazole gate near the solvent edge of the Fe-binding porphyrin ring [4], a medium-range interaction (a still smaller energy) that also plays a key role in discrimination between $O_2$ and CO binding [5]. While the separation of the proximate and distal His in Euclidean space on opposite sides of the porphyrin ring is small, their separation tangentially along globin chains looping around the porphyrin ring is large, and in Mb and Hemoglobin ($\alpha,\beta$) it is usually 29 aa. This fixed chain spacing is not so surprising for Mb (~ 75% aa conservation between human and chicken), or for Hb (~ 70% conservation), yet their functions (storage or transport) are different. Moreover, the recent discovery of Nb, which appears to resemble Mb [6], has shown a dis-his Nb spacing of 32 aa in all species sequenced so far. This could be related to a different oxidation channel opposite to the standard Mb channel, as shown by detailed structural comparisons ([7], esp. Figs. 2 and 3, and [8]). Here we are involved with still smaller energies.

The proximate - distal His channel aa spacing in Mb has oscillated from 32 (ultraprimitive vertebrates, lamprey and hogfish) to 25 (amphibian bullfrog), and finally settled at 29 in Mb and Hb $\alpha$ and $\beta$, across many species, including even primitive pufferfish Mb and lungfish Hb. The mechanics of channel opening and closing involve long-range interactions (network strain), not just medium range channel unblocking near His64. Only recently has it become possible to show the existence of a single Mb oxidation channel spanning so many aa side groups, and involving long diffusion times, with a single escape pathway, using Newtonian mechanical models [9-11].

**Methods**

For the reader's convenience we summarize here the main features of our recently developed bioinformatic scaling method [12]. The entire approach is post-Newtonian, and initially it may appear counter-intuitive. The central starting point is the implicit information contained in amino acid sequences in NCBI and their related structures in the PDB. Quite generally, one suspects that buried in these data are several key features of Nature's solutions for specific proteins to Levinthal's paradox (how proteins can function when they involve many exponentially complex, small energy differences) [13]. Newtonian all-atom simulation models that involve motions of thousands of atoms are not only extremely cpu intensive (typically months to reach a few results), but they do not address the evolutionary nature of protein sequence-structure-function relations, or provide tools for extracting functional information hidden in NCBI and PDB.

The first break in addressing this problem was the landmark discovery of a uniquely effective hydropathicity scale [14]. Competing effects of hydrophobic and hydrophilic segments of a given protein have long been known to be the primary driving force behind the folding of protein chains into protein globules. There are secondary effects associated with longitudinal hydrogen bonding (α helices) and transverse hydrogen bonding (β strands), and even weaker charge effects, but in most proteins the dominant physico-chemical factor in an overall property such as aggregation [12,15] is hydropathic interactions. This new scale focuses on the hydrogen bonds in the interface between the water film and the protein. It is based on the solvent-accessible areas of amino acid van der Waals surfaces partitioned in overlapping regions according to the Voronoi rules. It has fundamental value because it describes globular protein sequence-structure-function relations in the context of a thermodynamic model of structural transitions dominated by the rearrangement energies of the protein-water interfacial hydrogen bond network.

The second step is unexpected, but the justification of it has become apparent from studying many examples. The first popular application of hydropathicity scales $\Psi$ was to the study of all-α heptad transmembrane opsins, whose seven internal transmembrane segments are predominantly hydrophobic, with a typical length around 20 amino acids. One can carry this description to the similarly 20 amino acid thick layers adjacent to cell membranes where proteins can interact most effectively, as they are temporarily confined to a narrower space. Some readers will recognize a similarity to heterogeneous catalysis, with the membrane here serving as catalytic substrate. This picture is a generalization of enzyme catalysis (proteins with organic molecules) to all protein-protein interactions and evolution in vivo. It also helps to resolve Levinthal's paradox [13] on a functional level wider than protein folding, a larger puzzle than the big bang of cosmology.

In this membrane-based surface space proteins can both interact on a 20 amino acid length scale, or evolve via exchange of modular elements of ~20 amino acids. This is a post facto explanation for the remarkable discovery made by [14]. Instead of restricting the calculation of solvent-accessible amino acid areas to entire folded proteins from the PDB, [14] studied those areas associated with protein fragments, centered on specific amino acids, of length L, $9 \leq L \leq 35$, an interval itself centered on the typical membrane thickness 21. (These fragments are sometimes called modules in models of protein evolution [16].) These solvent-accessible areas were then found to be linear in a log-log plot against L, over the same range, for all 20 amino acids. This collinearity is indicative of special properties of self-organized (nearly perfect) protein networks called self-similarity, and has been conjectured to be the key to a wide range of other nearly perfect networks. The amino-acid specific parameters of the model are universal fractals [17, 18].

Our hydropathic analysis shows that small-gas ligand ($O_2$, CO, NO,…) entry and exit globin allosteric dynamics can be dominated by longitudinal hydropathic strain at a single chain length, the spacing between proximate receptor - distal gate His. Compared to enriched multi-dimensional Newtonian all-atom simulation models [9-11], our hypermodern approach based only on one-dimensional optimized hydropathic amino acid square wave profiles may appear simplistic. However, there are compensating advantages, as the economical methods used here have been applied to many other proteins with many different functions [19,20]. As we will see, evolution has employed similar optimizing tools on many proteins, which are recognizable from the very large amino acid sequence data base alone, with minimal support from known secondary α helix and β strand structural data.

The motivation for constructing allosteric hydropathic profiles is also based on thermodynamic concepts concerning the differential aqueous sculpting of globular protein surfaces near a critical point [12]. It can be compared to elastometric treatments of hinge-bending conformational transition pathways [21-23]. Euclidean structural models are static, but some dynamical features can be inferred by mutagenesis experiments that replace larger aa by smaller aa [8]. Note that mutagenic experiments involve non-WT proteins, whereas our bioinformatic scale involves only the differential surface geometry of WT proteins, a much smaller space with strong functional constraints.

Because globins perform the key elementary biochemical metabolic functions and are very well studied for many species, one can use them to extend and refine some previous correlations between hydroprofile extrema and species functionality, especially in connection with rapid escape from predators. Previously these largely referred to leveling of extrema as a mechanism to accelerate conformational motions by synchronizing chain segments that are well separated along the chain, but may be close in Euclidean space because of chain folding [19]. Here evolution has conserved the distal-proximate His spacing over many species, so that the same constructs can be used repeatedly, producing results that are nearly parameter-free and strongly over-determined.

**Results: Two Channels**

Mb, Hb α and β, and Nb globin sequence identities (positives) are less than 25(40)%. [7] presents a structure-based sequence alignment showing 75% α helices and heme contact points (their Fig. 1), as well as views of superposed Mb and Nb numbered structures (their Fig. 2(A)). Apparently the heme stabilizes a common Mb-Nb globin fold, while the oxidation channels of Mb and Nb differ. These superposed Euclidean structures [7] can be compared to the human Mb and Nb hydroprofiles shown here in Fig. 1.

The Mb channel, which exits past the distal His E7 gate, is opposite to the Nb "apolar" tunnel, which is analogous to the exit past His100 in a mini-hemoglobin, as shown in Figs 1A and B of [8]. Experimental data, as well as the recent "milestone" calculation for Nb [9], agree that each of these opposing channels is unique. Evolutionary optimization should favor a more efficient single channel, and our hydropathic results in Fig. 1 show a simple elastic explanation for the existence of two opposing channels in Mb and Nb. Relative to Mb, the Nb profile shows a small softening near the at the beginning of the G chain, but the largest change by far is the Mb opening of a deep hydrophilic elastic hinge near 50, which releases the His E7 (64) imidazole ring gate for an exit channel.

Stated differently, the Nb hinge near 50 is much less hydrophilic than the deep Mb valley. The soft Mb hinge is centered on the 17 aa sequence 45DKFKHLKSEDEMKASED61, which is strongly hydrophilic, as 13aa are K, D or E. Using our rescaled version [15] of the MZ bioinformatic scale [20], which has an average hydropathicity ~ 155, the average for the soft Mb 45-61 sequence is 115 (strongly hydrophilic), compared to 45N**C**RQFSSPED**C**LSSPE60 = 132 (only moderately hydrophilic; only 4aa are R,D or E) for Nb. These elastic sequences fill the chains between the mainly hydrophobic heme contact points (42,44) and (64,67,68,71) [7]. The full range of the contact points is again W = 71-42 = 29! According to Hooke's law (1660, 1678), the average elasticity of this range is a rough estimate of the extent to which the distal His E7 imidazole ring gate (64) can open to create a channel entrance and exit.

Overall the human hydroprofiles shown in Fig. 1 contain three parts, hydrophobic near the N- and C- terminals, and an active central part 50-110 centered between distal and proximate His. In addition to the soft hinge near 50 the central region in Mb is bounded by a weaker second hinge above the proximate His F8 (96). The overall globular structure is stabilized by weakly hydrophobic shoulders near the C- and N- terminals. In Nb, the softest region has switched to be centered on the proximate His F8 (96), and it seems very likely that this is another indication that the exit channel of Nb is different from that of Mb.

The largest evolutionary changes occur between aquatic and terrestrial species, so Fig. 2 shows Mb and Nb hydroprofiles of zebrafish, a tropical freshwater fish. It is immediately obvious that the largest difference is the upward (hydrophobic) shift of the Mb hinge near 50 from human to zebrafish. The Nb profile is little changed from fish to human in the 50-110 central region: for example, the correlation there of (human, zfish) for $\psi(aa, W = 29)$ is 0.62, while the same correlation for Mb is 0.41. Nb (Zfish) exhibits a large number of nearly level hydrophobic extrema, and the main change in Nb (human) is a small hydrophilic shift of the active central region, presumably increasing its activity.

Hydroprofiles are sensitive to function and environment, even for Nb in freshwater fish, for example, zebrafish (tropical) compared to trout (temperate). As shown in Fig. 3, the N- and C-terminal hydrophobic stabilization trout shoulders (buttresses) are increased, so that the hydrophobic extrema change from flat (Zfish) to parabolic (trout). Meanwhile the central region, involving the oxidation channel, is almost unchanged.

**Neuroglobin Evolution**

There are key changes from aquatic to terrestrial Nb, specifically trout to chicken (Fig. 4). The ordering of the hydrophilic extrema reverses, but only slightly, which is probably not enough to shift the exit channel to distal His 64 in chicken. A new terrestrial feature emerges, as the hydrophobic maxima develop a W pattern, with the new central peak near 80. This is the Euclidean apex of the chain segments leading to distal His 64 and proximate His 96 on opposite heme sides [7]. Stiffening the central apex stabilizes the network contacting the heme. Studies of escape channels marked by pockets that accommodate Xe suggest that the apical chain region halfway between distal His 64 and proximate His 96 is the only region common to Mb and Nb small molecule channels [24]. It is possible that mouse Nb uses both distal His 64 and proximate His 96 channels for faster oxidation.

The chief Nb change from chicken to mouse (Fig. 5) is a large mouse hydrophilic shift (35% of the profile width near the center hydrophobic maximum, and 40% at the proximate His F8 hydrophilic minimum). Making the overall Nb profile more hydrophilic enables mouse to act more quickly, which is appropriate for a small rodent prey. Note the leveling of the two hydrophilic extrema. This facilitates synchronization of conformational transitions (Fig. 3 of [19]), desirable for immediate mouse response to danger. The visual systems of primitive chickens were already optimized for perceiving rapid movement [25].

In earlier studies [12] of hydroprofiles based on the bioinformatic MZ scale [14,15], it was noted that because the MZ scale has both a thermodynamic and an evolutionary basis in terms of self-organized criticality, its results are generally more informative than those obtained with the standard KD scale, based on water-air complete chain unfolding [26]. Because the large differences between Nb chicken and mouse shown in Fig. 5 with the MZ scale are so informative, Fig. 6 shows the same profiles calculated with the KD scale. Mouse is still more hydrophilic, but most of the differences have decreased sharply: for instance, the 40% MZ difference at the proximate His F8 hydrophilic minimum is < 10% with the KD scale. Also the KD hydrophilic extrema are not level, which disrupts conformational synchronization. The improvements using the MZ scale compared to the KD scale are an internal consistency check on the relevance of hydroprofile extrema.

Rabbit is much larger than mouse, and can afford a more sophisticated delayed escape strategy [27]. It also lives more than twice as long, and benefits from a more hydrophobic Nb profile, which is what Fig. 7 shows. Humans are much larger and live much longer, but the differences between mammals are small (approaching optimized criticality), see Fig. 8. In the proximate His 96 region Human Nb closely resembles Mouse Nb, while Human is more hydrophobic and more stable in the N- and C-terminal shoulders, consistent with greater longevity.

In most species' Nb, there is a Cys disulfide bond between the C and D helices, with a variable Cys-Cys spacing of 6-9 aa (for example, human 45N**C**RQFSSPED**C**LSSPE60). This disulfide bond must regulate dynamics of the CD loop [28]. While the disulfide bond is present in trout, chicken, rabbit and human, it is strikingly absent in mouse, where the dis-prox His tangential chain spacing is fully conserved. This is consistent with the extremally hydrophilic Nb profile of mouse in the central region between dis and prox His, consistent with the mouse's need for rapid escape.

**Myoglobin Evolution**

Mb is one of the most studied proteins [1,3,4], and the emerging evidence, particularly from fish species, indicates that Mb fulfills a broad array of physiological functions in a wider range of different tissues [29]. However, almost all Mb structural studies have focused on sperm whale or human Mb. Studies of tuna have shown that the D helix centered on 55 has been disordered [30]. As we saw in discussing Fig. 1, the key to the flexibility of the His 64 gate is the hydrophilic softness of the chain region between the heme contact points (42,44) and (64,67,68,71) [7]. Helix D stabilizes this chain region, and when this region becomes hydroneutral (Fig. 9), it is possible that in the absence of helix D the distal His E7 may still function as a gate for a channel similar to the terrestrial oxidation channel, but probably more complex. At present the Mb aquatic oxidation channel is unknown.

For mouse we concluded that Nb is especially hydrophilic because mouse must recognize danger quickly (Nb is concentrated in retina and brain [31]). The most striking new feature, after chicken, of Mb in mouse (Fig. 10) is the appearance of the apical hydrophobic peak at 80, which we suggested above stabilizes the network contacting both sides of the heme. Such stabilization could increase signal transmission and release oxygen more rapidly to tissues.

The final stage of Mb evolution from mouse to human is shown in Fig. 11. As with Nb, mouse is an extremum between chicken and mammals, especially near the apical site 80 heme stabilizing hydrophobic peak. The overall differences are smaller with Mb than with Nb, which was to be expected, as Nb is finely tuned to function optimally in neural networks, whereas Mb has many functions in all tissues.

We saw in Figs. 5-7 how the extremal properties of mouse Nb appeared as much larger differences with the MZ conformational (second-order) scale, than with the KD scale. In Fig. 12 we show the Mb profiles of Fig. 11 (MZ scale), now with the KD scale. With Mb profiles obtained with the two scales are much closer, and the differences are now about 10% larger with the KD scale. At first this seems surprising, but it could be merely a reflection of the fine tuning of Nb for neural networks, and the wide range of functions of Mb, which are better described using the first-order KD scale suited to protein unfolding. In the network glass and polymer literature these heterogeneous first-order effects [32] are sometimes described as random, which is simplistic [33]. The homogeneous limit of dynamical network stretched exponential relaxation is far from random, and involves topological fractals similar to the MZ scaling fractals [34].

**Discussion**

Both the structure and function of a protein chain in its native state are determined by its amino acid sequence [35]. Mutagenic and structural studies have established many Euclidean short-range features of Mb and Nb, including the existence of different oxidation channels [1,4, 7-11].

Hydropathic scaling provides a different and complimentary picture, involving long-range (allometric) elastic interactions at the molecular level, which connects amino acid sequences directly to function, as well as elucidating previously hidden structural features. The most striking feature of all Mb and Nb profilesis their tripartite structure, which is barely visible in aquatic profiles (especially tropical), but is well developed and closely related to function in terrestrial species. The central peak is associated with the porphyrin ring, while the side extrema facilitate allometric synchronized motion of N and C wings. This produces symmetric patterns, especially for mouse, which it requires for its escape from predators.On a more detailed level, profiles identify the elastically flexible region near 50 in Mb, which enables His E7 imidazole ring to function as the Perutz gate in Mb. In Nb the hydrophilic region near 50 stiffen and it is no longer the minimum. That minimum has shifted to near the proximate His at 96, and this shift is accompanied by the appearance of an oxidation channel on the opposite side of the heme. Thus tripartitie hydropathic profiles alone strongly suggest different escape channels, from opposite sides of the heme, in Mb and Nb.

The extreme elasticity of the Mb hinge near 50 should be compared to that of several more elastic regions: in the 230-260 region of the A4 (amyloid precursor) 770 aa protein of the neural network, and near the center of 413 aa Aspein, which is responsible for directed formation of calcite in the shell of the pearl oyster – an ultrasoft buffer supporting the growing ultrahard oyster shell [36,37].

Returning to Figs. 1 and 2, one notices that Mb has evolved from aquatic to terrestrial species much more than Nb. At first, this seems surprising, but aquatic (terrestrial) species are cold- (warm-) blooded, and ligand exchange kinetics in respiratory tissues of aquatic species may be quite different from that in terrestrial species [38]. By contrast, the neural and retinal networks of all vertebrates differ largely in size [36].

The leveling of hydropathic extrema emphasized in Figs. 3-8 has been discussed by differential geometers in terms of level sets as applied to the diffusive flow of liquids, in at least 700 papers

[39,40]. The synchronization comparisons made here are driven by primary biological forces (prey vs. predator).

**Conclusions**

The systematics of Mb and Nb evolution can be accurately described using two hydropathic scales [14,26], with the second-order MZ scale being much more accurate for Nb, and the first-order KD scale being slightly more accurate for Mb. The evolution of the competing enzymatic and lytic functions of lysozyme *c*, and its role in suppressing amyloid aggregation, are described much more accurately by the MZ scale [12]. This scale represents the quantitative realization, using the tools of thermodynamics and differential geometry, of the general theory of dominant conformational motions slaved by the hydration shell and the bulk solvent [41]. Specific soft modes are seen in Mb-$O_2$ compared with those in Mb-CO [42].

It is interesting to compare the present molecular scaling treatment of the evolution of allometric interactions with macroscopic models relating vertebrate metabolism to surface area [43]. It appears that these vary significantly between fish, amphibians, reptiles, birds and mammals [44]. There is also substantial scatter in visual response rates when these are scaled by size or metabolism [45]. Here we have seen significant differences even between freshwater and ocean fish (Fig. 3). We have easily connected differences between mouse Nb and other species Nb to the mouse's predator escape strategy. Scaling, improved by inclusion of multiple individual factors, is likely to be an increasingly important tool in medical contexts [46,47]. At the molecular level it has already provided a detailed description of viral sequence evolution [48], which may well be more accurate in selecting suitable vaccine targets than antigenic clustering [49]. Because of the complexity of proteins and their interactions, economy, simplicity and universality are critical features of theoretical biophysical tools.

**Methods** Although bioinformatic scaling is novel, the calculations described here are very simple, and are most easily done on an EXCEL macro, with sequences from NCBI. The one used in this paper was built by Niels Voohoeve and refined by Douglass C. Allan.

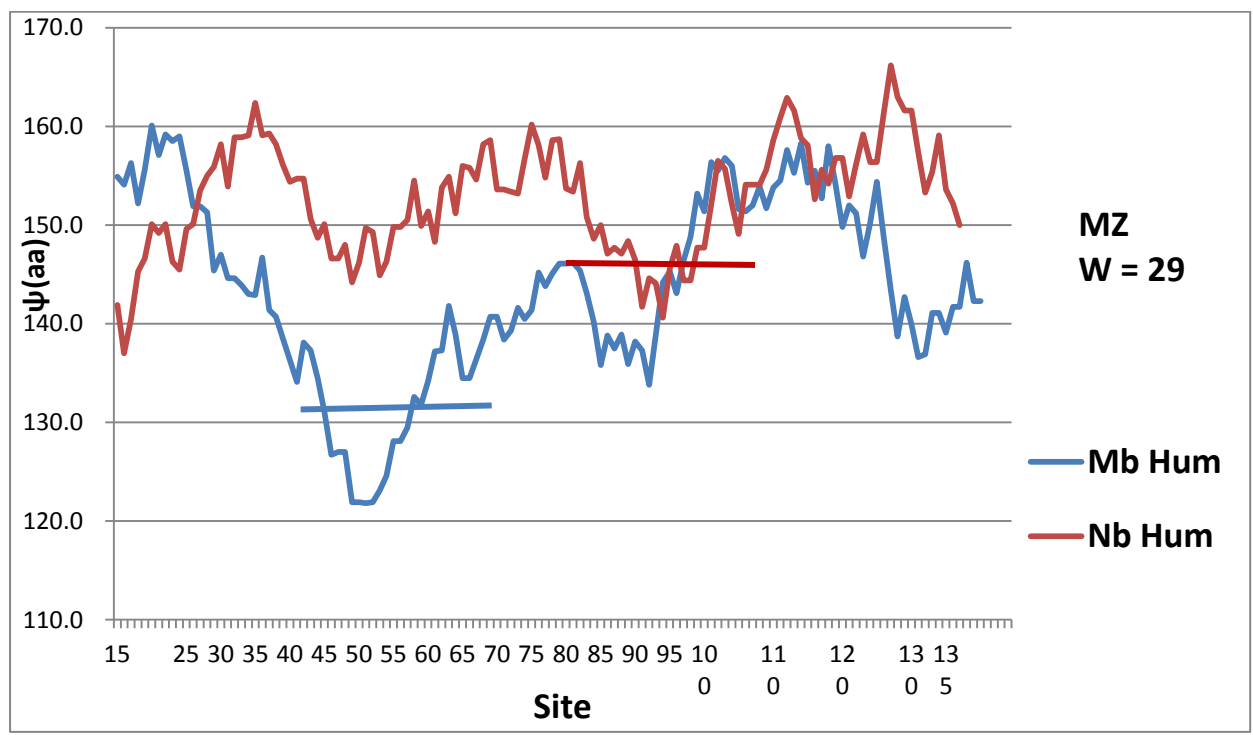

Fig. 1. The $\Psi(aa, W=29)$ hydropathic profiles for human Myoglobin and Neuroglobin. Also shown is the $\psi(aa)$ averages for the elastic region spanning 29 amino acids between the Mb heme contact points 42-71 discussed in the text. The much lower elastic blue value for Mb enables the E8 His (64) gate to open and function as a small gas molecule channel. The much larger red elastic value for Nb stiffens this region, so that the functional channel is switched from Mb. In Nb the channel goes from E8 (64) to the C-terminal ends of the E and H helices [7,8]. Note that in Mb the deepest hydrophilic minimum is centered near the distal His 64, while in Nb the deepest minimum is centered near the proximate His 96.

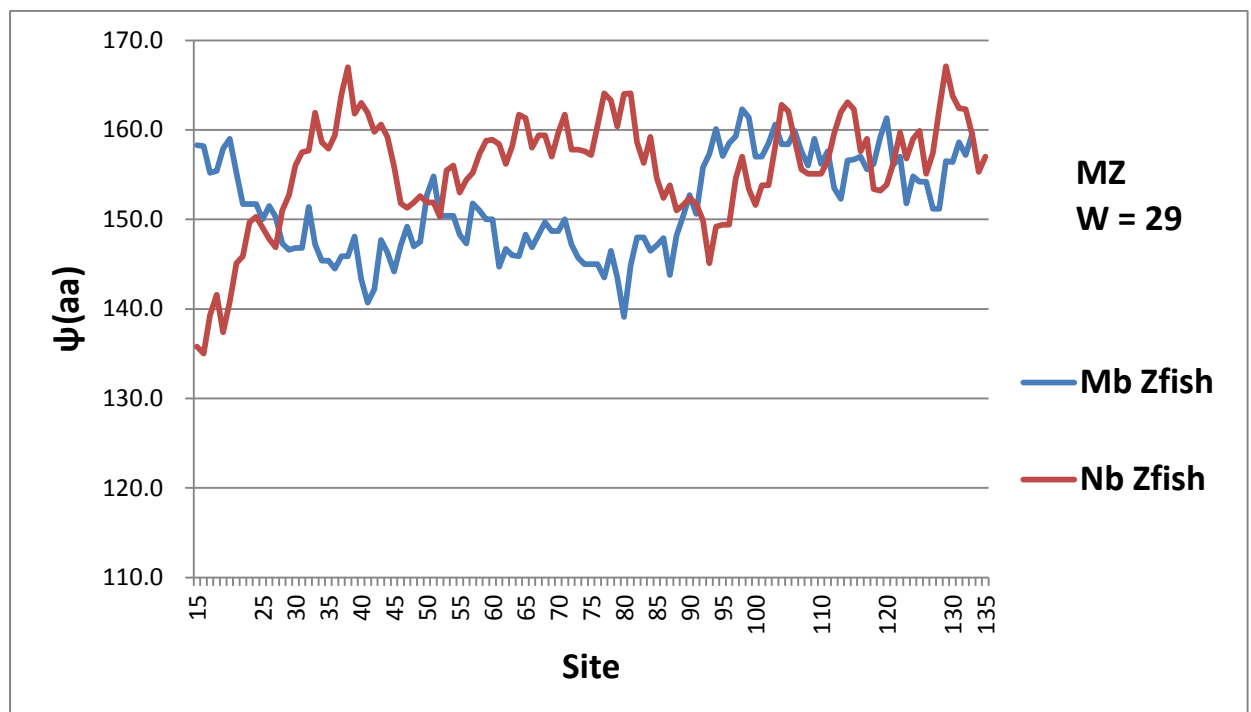

Fig. 2. The zebrafish profiles here should be compared to the human profiles of Fig. 1. The largest difference is the upward (hydrophobic) shift of the Mb hinge near 50 from human to zebrafish. The Nb profile is little changed from fish to human in the 50-110 central region, which is shifted downwards in human by only 4 (about 10% of the overall range) in $\Psi(aa, W = 29)$.

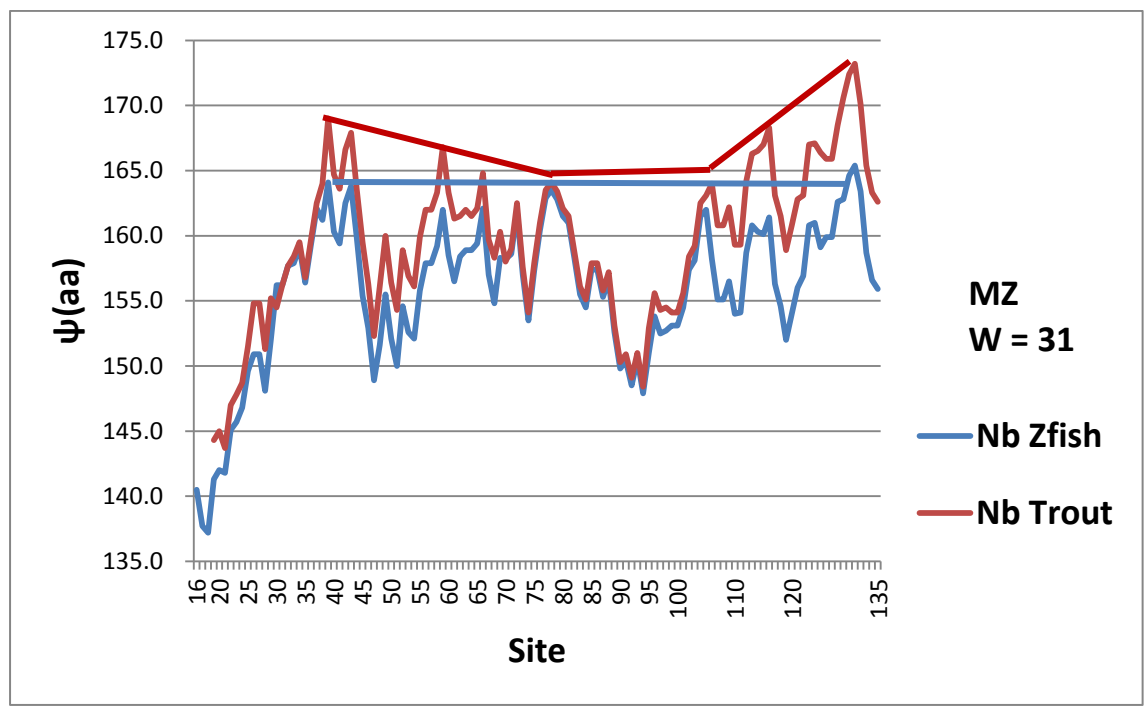

Fig. 3. Comparison of the hydroprofiles of tropical and temperate freshwater fish. The central region involving the heme and the distal His 64 – proximate His 96 channel is conserved, while the remaining regions are parabolically more hydrophobic for temperate trout, and nearly level for tropical oceanic Zebrafish.

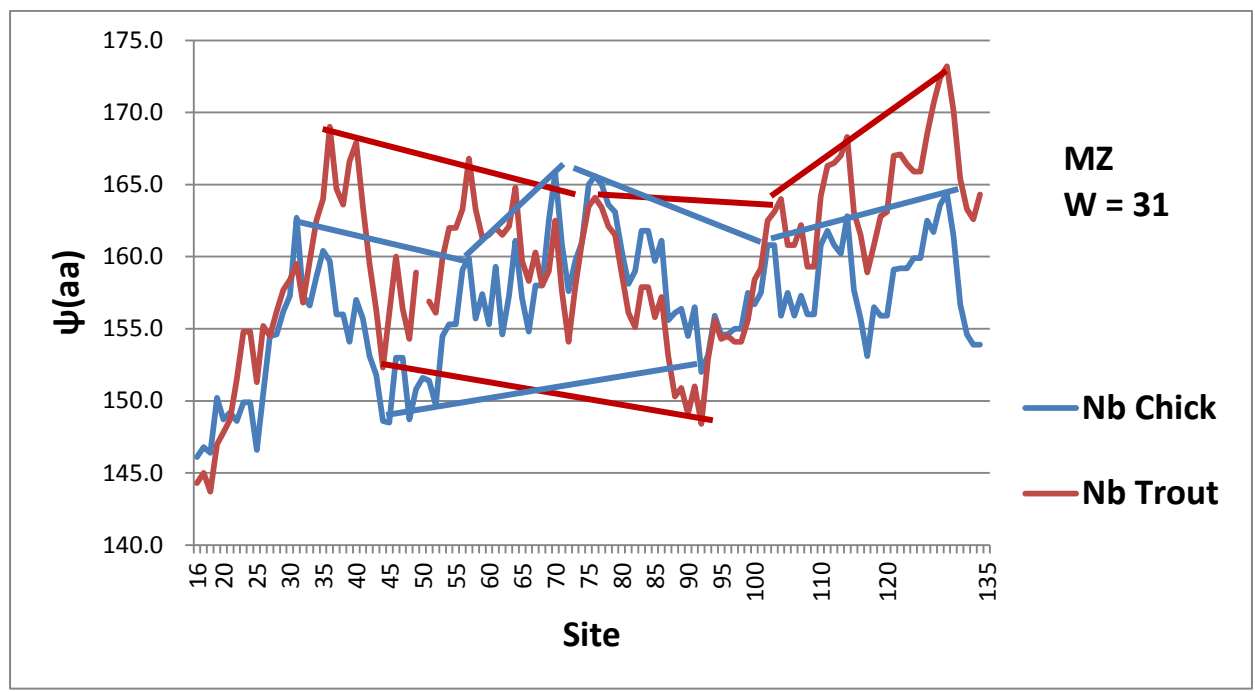

Fig. 4. Nb chicken is unusual (see text).

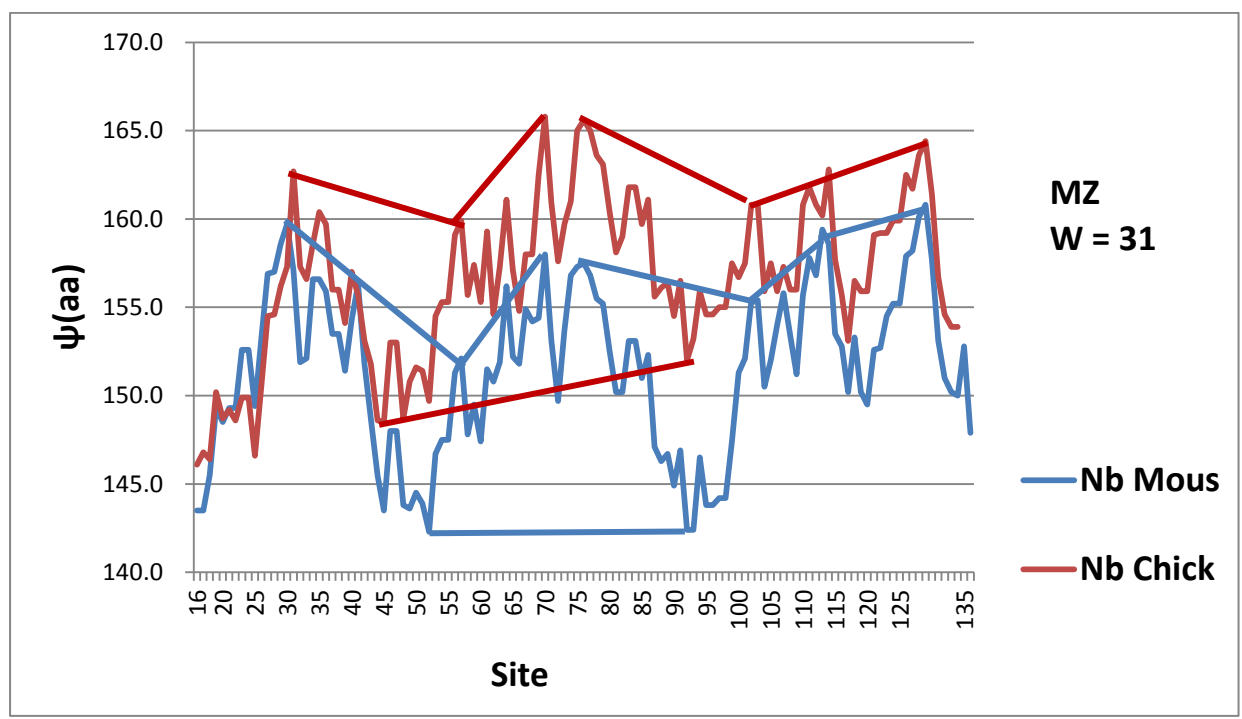

Fig. 5. Comparison of Nb for chicken and mouse with the MZ scale.

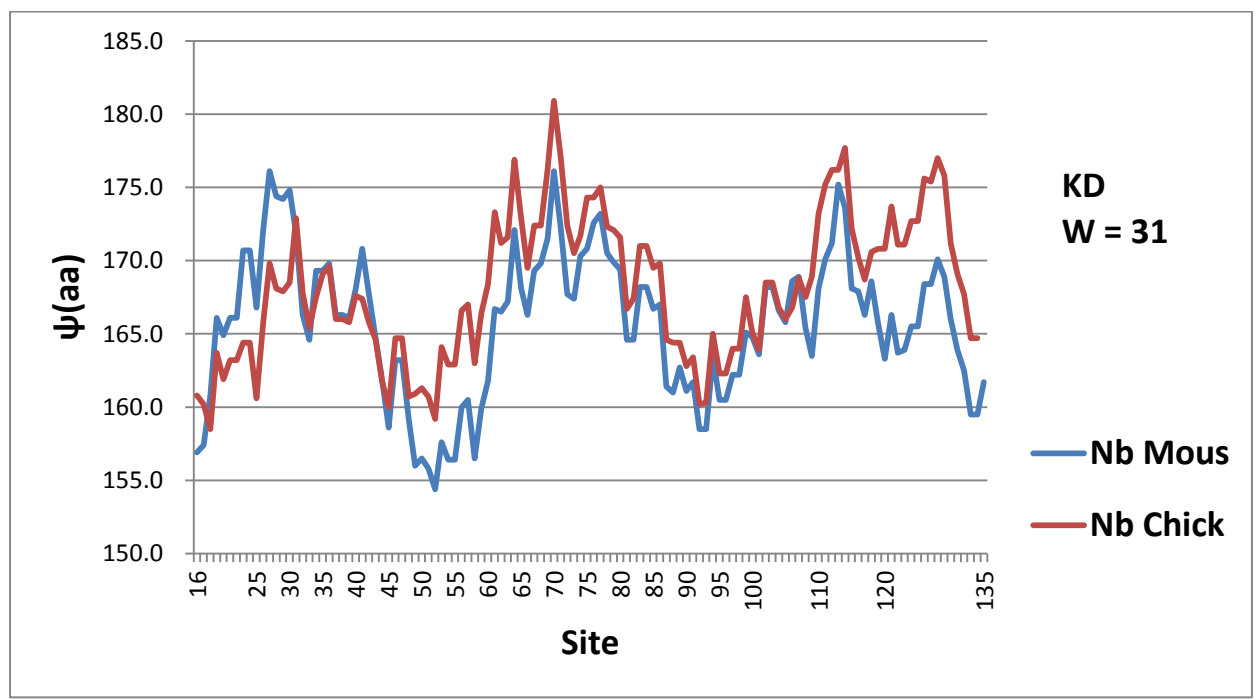

Fig. 6. Comparison of Nb for chicken and mouse with the KD scale. The large differences seen with the MZ scale in Fig. 5 have become small or disappeared.

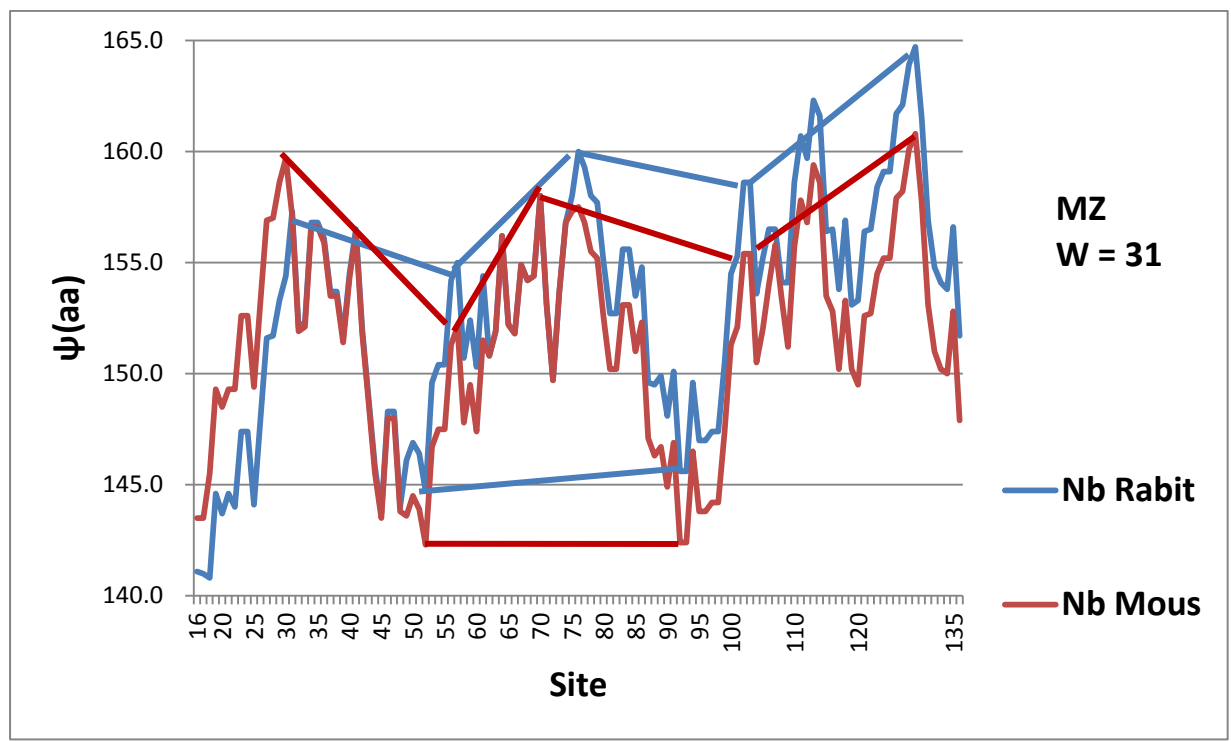

Fig. 7. Rabbit Nb is more hydrophobic than mouse Nb, and its hydrophilic hinges are less nearly level.

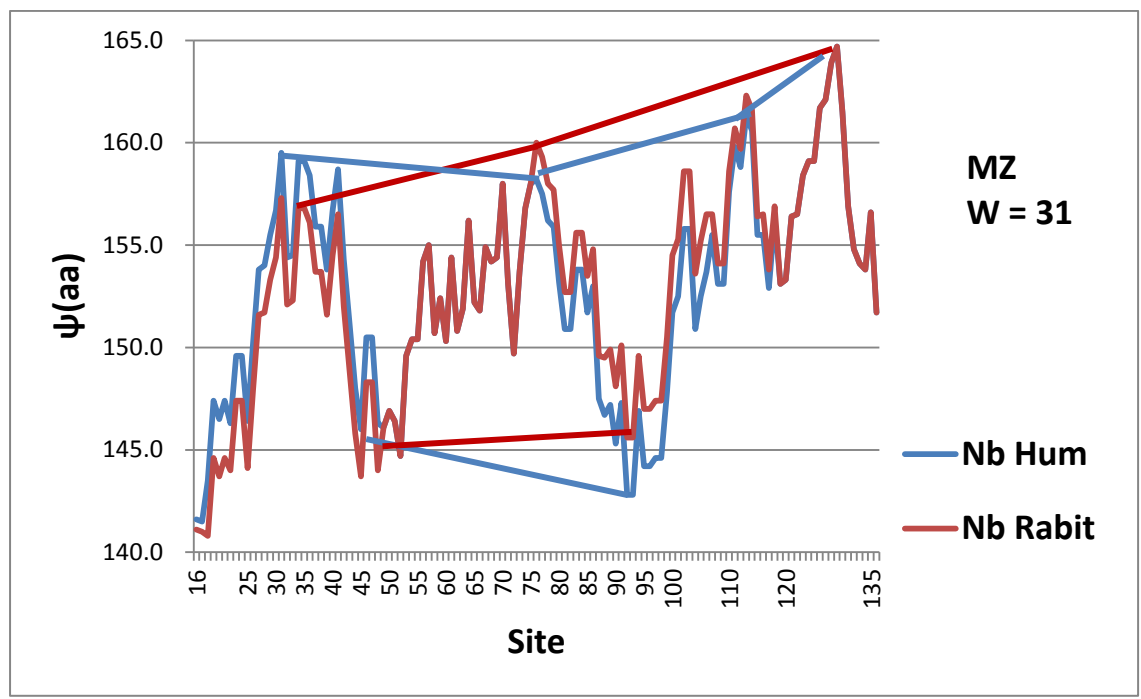

Fig. 8. Compared to rabbit Nb, human Nb is stabilized near the N terminal, and more hydrophilic at the proximate His 96 site.

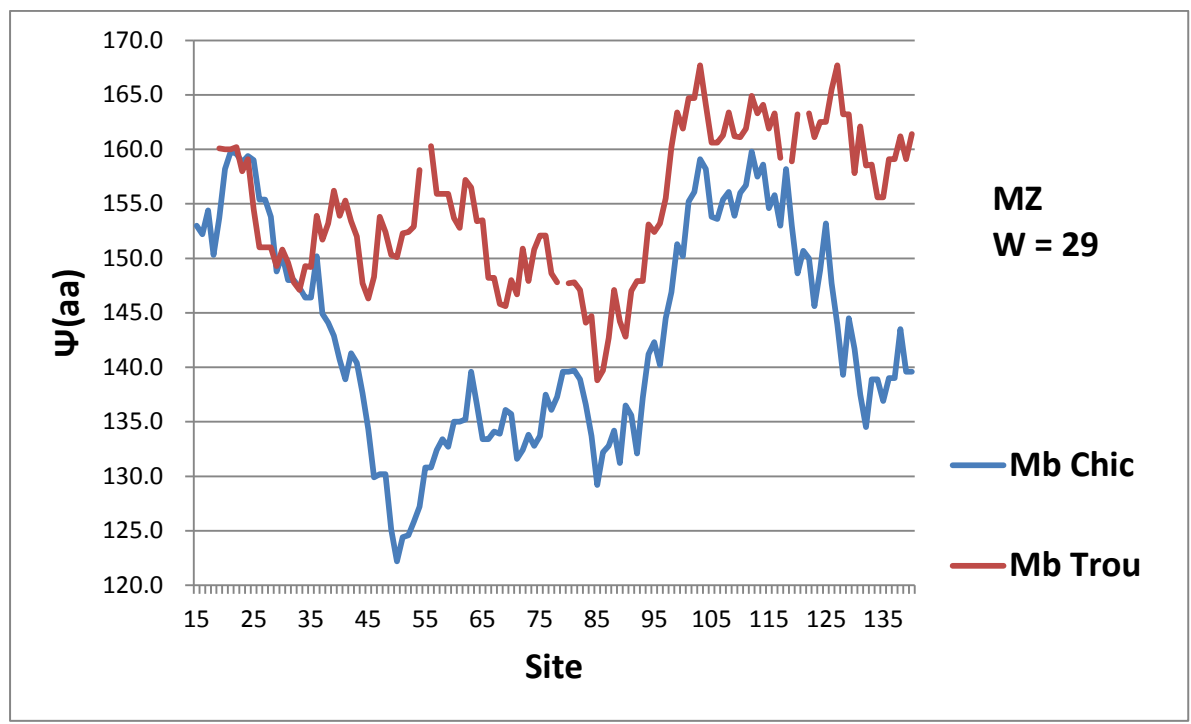

Fig. 9.  Hydroprofiles of Mb trout and chicken.  Trout is much more hydrophobic, and the elastically soft hydrophilic region near 50 has become hydroneutral.

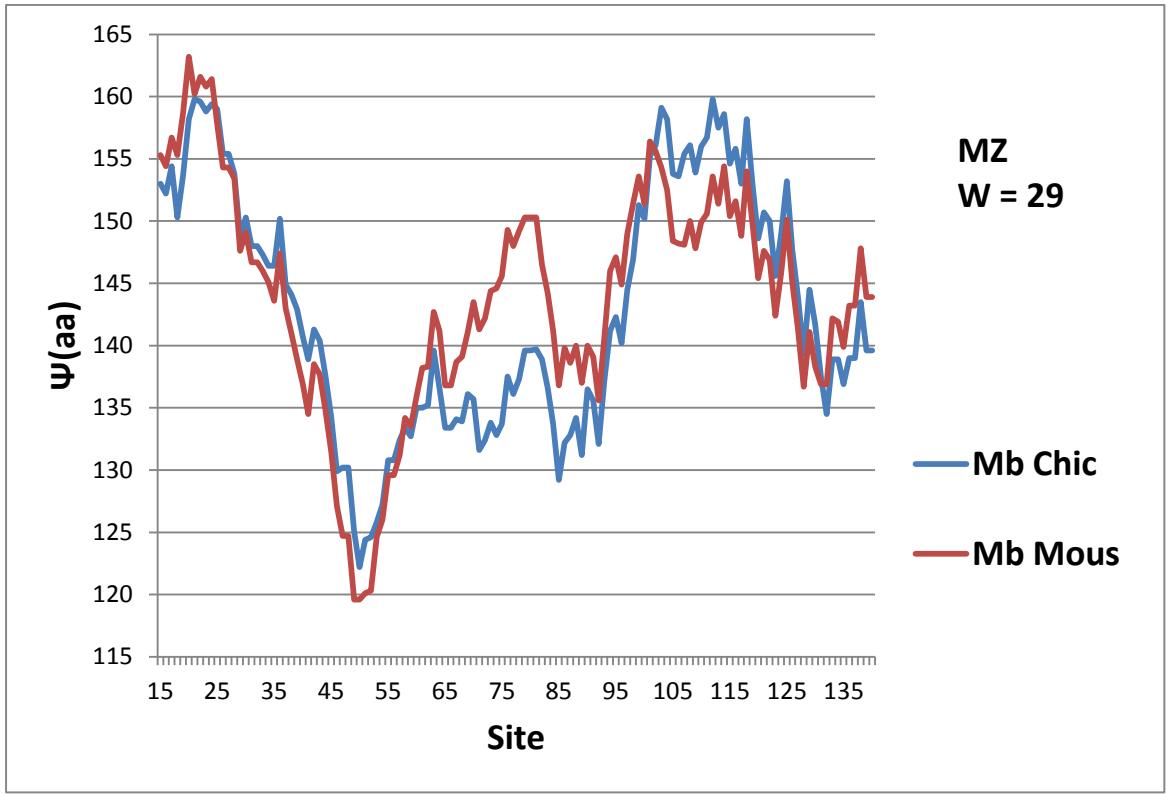

Fig. 10. The largest change here is the mouse apical hydroneutral peak near site 80.

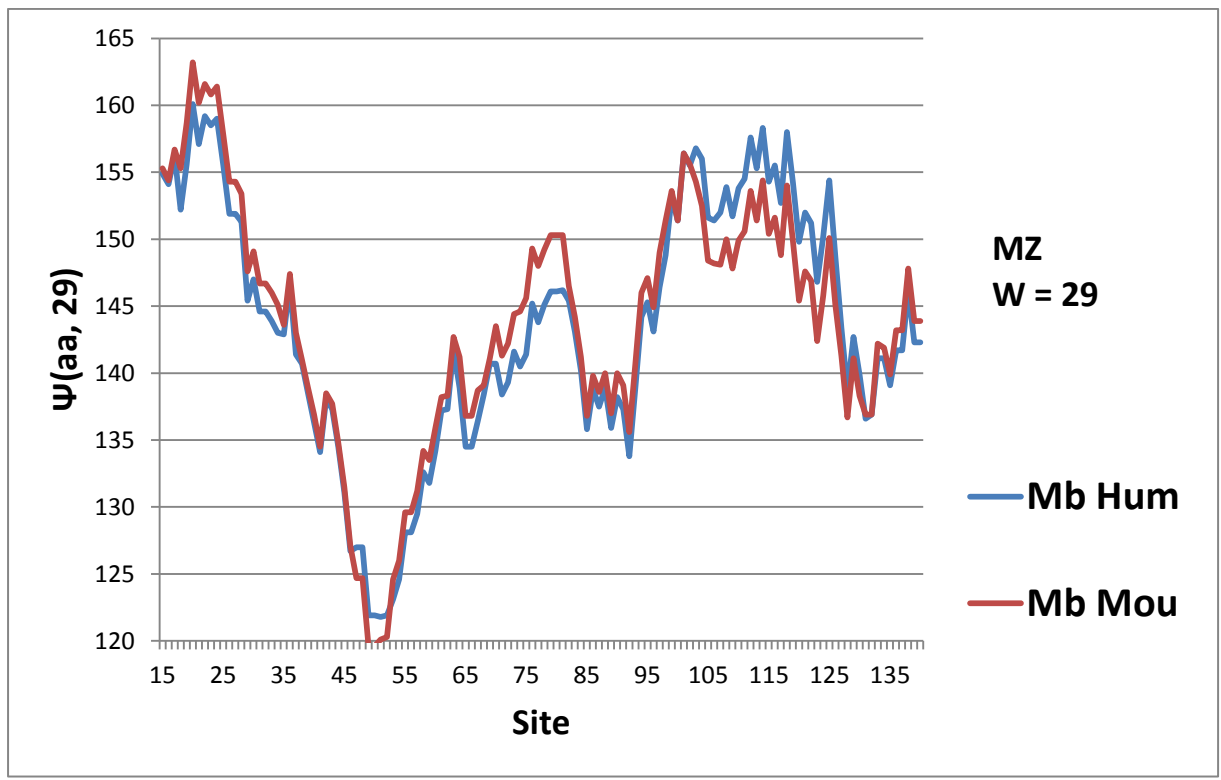

Fig. 11. Mb has stabilized from rodents to mammals.

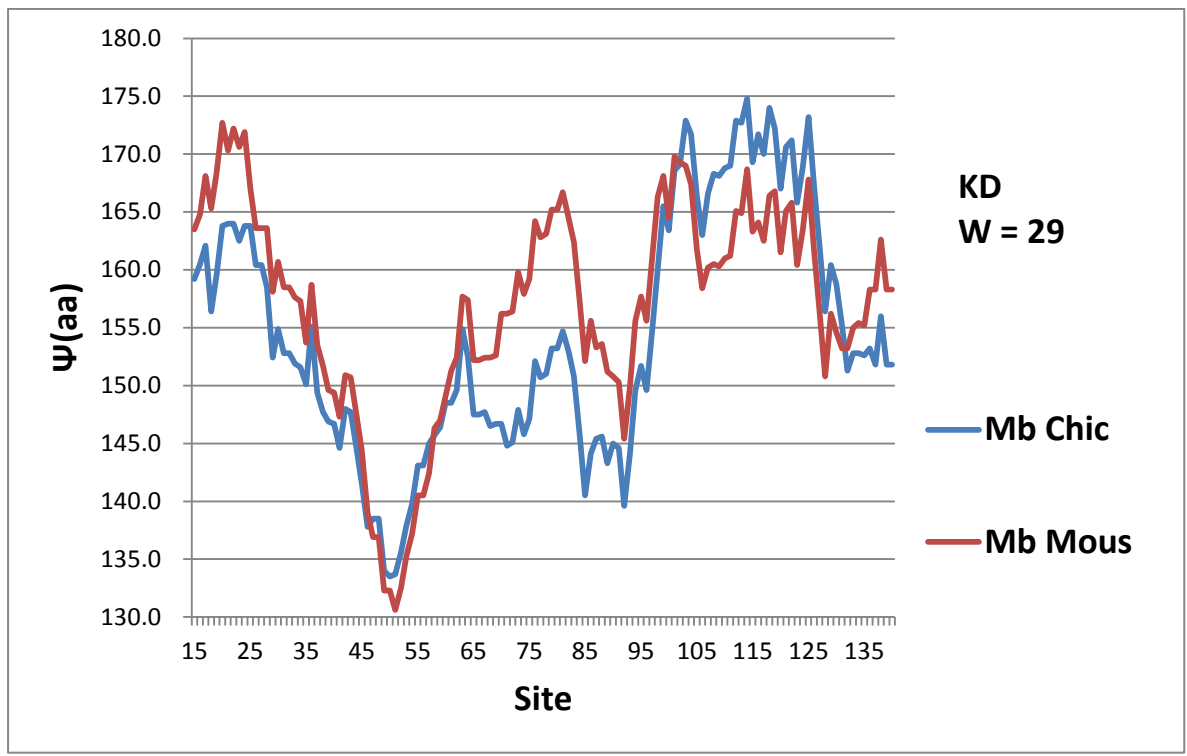

Fig. 12. Here the profiles are evaluated using the first-order KD scale, to be compared with Figs.10, 5 and 6.